\documentstyle[twoside,epsf]{article}

\catcode`\@=11
\long\def\@makefntext#1{
\protect\noindent \hbox to 3.2pt {\hskip-.9pt  
$^{{\eightrm\@thefnmark}}$\hfil}#1\hfill}		%CAN BE USED 

\def\thefootnote{\fnsymbol{footnote}}
\def\@makefnmark{\hbox to 0pt{$^{\@thefnmark}$\hss}}	%ORIGINAL 
	
\def\ps@myheadings{\let\@mkboth\@gobbletwo
\def\@oddhead{\hbox{}
\rightmark\hfil\eightrm\thepage}   
\def\@oddfoot{}\def\@evenhead{\eightrm\thepage\hfil
\leftmark\hbox{}}\def\@evenfoot{}
\def\sectionmark##1{}\def\subsectionmark##1{}}

\oddsidemargin=\evensidemargin
\renewcommand{\thefootnote}{\fnsymbol{footnote}}

\newcounter{sectionc}\newcounter{subsectionc}\newcounter{subsubsectionc}
\renewcommand{\section}[1] {\vspace{12pt}\addtocounter{sectionc}{1} 
\setcounter{subsectionc}{0}\setcounter{subsubsectionc}{0}\noindent 
	{\tenbf\thesectionc. #1}\par\vspace{5pt}}
\renewcommand{\subsection}[1] {\vspace{12pt}\addtocounter{subsectionc}{1} 
	\setcounter{subsubsectionc}{0}\noindent 
	{\bf\thesectionc.\thesubsectionc. {\kern1pt \bfit #1}}\par\vspace{5pt}}
\renewcommand{\subsubsection}[1] {\vspace{12pt}\addtocounter{subsubsectionc}{1}
	\noindent{\tenrm\thesectionc.\thesubsectionc.\thesubsubsectionc.
	{\kern1pt \tenit #1}}\par\vspace{5pt}}
\newcommand{\nonumsection}[1] {\vspace{12pt}\noindent{\tenbf #1}
	\par\vspace{5pt}}

\newcounter{appendixc}
\newcounter{subappendixc}[appendixc]
\newcounter{subsubappendixc}[subappendixc]
\renewcommand{\thesubappendixc}{\Alph{appendixc}.\arabic{subappendixc}}
\renewcommand{\thesubsubappendixc}
	{\Alph{appendixc}.\arabic{subappendixc}.\arabic{subsubappendixc}}

\renewcommand{\appendix}[1] {\vspace{12pt}
        \refstepcounter{appendixc}
        \setcounter{figure}{0}
        \setcounter{table}{0}
        \setcounter{lemma}{0}
        \setcounter{theorem}{0}
        \setcounter{corollary}{0}
        \setcounter{definition}{0}
        \setcounter{equation}{0}
        \renewcommand{\thefigure}{\Alph{appendixc}.\arabic{figure}}
        \renewcommand{\thetable}{\Alph{appendixc}.\arabic{table}}
        \renewcommand{\theappendixc}{\Alph{appendixc}}
        \renewcommand{\thelemma}{\Alph{appendixc}.\arabic{lemma}}
        \renewcommand{\thetheorem}{\Alph{appendixc}.\arabic{theorem}}
        \renewcommand{\thedefinition}{\Alph{appendixc}.\arabic{definition}}
        \renewcommand{\thecorollary}{\Alph{appendixc}.\arabic{corollary}}
        \renewcommand{\theequation}{\Alph{appendixc}.\arabic{equation}}
        \noindent{\tenbf Appendix \theappendixc #1}\par\vspace{5pt}}
\newcommand{\subappendix}[1] {\vspace{12pt}
        \refstepcounter{subappendixc}
        \noindent{\bf Appendix \thesubappendixc. {\kern1pt \bfit #1}}
	\par\vspace{5pt}}
\newcommand{\subsubappendix}[1] {\vspace{12pt}
        \refstepcounter{subsubappendixc}
        \noindent{\rm Appendix \thesubsubappendixc. {\kern1pt \tenit #1}}
	\par\vspace{5pt}}

\newcommand{\textlineskip}{\baselineskip=13pt}
\newcommand{\smalllineskip}{\baselineskip=10pt}

\def\eightcirc{
\begin{picture}(0,0)
\put(4.4,1.8){\circle{6.5}}
\end{picture}}
\def\eightcopyright{\eightcirc\kern2.7pt\hbox{\eightrm c}}

\newcommand{\pub}[1]{{\begin{center}\footnotesize\smalllineskip 
	Received #1\\
	\end{center}
	}}

\def\abstracts#1#2#3{{
	\centering{\begin{minipage}{4.5in}\baselineskip=10pt\footnotesize
	\parindent=0pt #1\par 
	\parindent=15pt #2\par
	\parindent=15pt #3
	\end{minipage}}\par}} 

\renewenvironment{thebibliography}[1]
	{\frenchspacing
	 \ninerm\baselineskip=11pt
	 \begin{list}{\arabic{enumi}.}
	{\usecounter{enumi}\setlength{\parsep}{0pt}
	 \setlength{\leftmargin 12.7pt}{\rightmargin 0pt} %FOR 1--9 ITEMS
	 \setlength{\itemsep}{0pt} \settowidth
	{\labelwidth}{#1.}\sloppy}}{\end{list}}

\newcounter{itemlistc}
\newcounter{romanlistc}
\newcounter{alphlistc}
\newcounter{arabiclistc}

\newcommand{\fcaption}[1]{
        \refstepcounter{figure}
        \setbox\@tempboxa = \hbox{\footnotesize Fig.~\thefigure. #1}
        \ifdim \wd\@tempboxa > 5in
           {\begin{center}
        \parbox{5in}{\footnotesize\smalllineskip Fig.~\thefigure. #1}
            \end{center}}
        \else
             {\begin{center}
             {\footnotesize Fig.~\thefigure. #1}
              \end{center}}
        \fi}

\newcommand{\tcaption}[1]{
        \refstepcounter{table}
        \setbox\@tempboxa = \hbox{\footnotesize Table~\thetable. #1}
        \ifdim \wd\@tempboxa > 5in
           {\begin{center}
        \parbox{5in}{\footnotesize\smalllineskip Table~\thetable. #1}
            \end{center}}
        \else
             {\begin{center}
             {\footnotesize Table~\thetable. #1}
              \end{center}}
        \fi}

\def\@citex[#1]#2{\if@filesw\immediate\write\@auxout
	{\string\citation{#2}}\fi
\def\@citea{}\@cite{\@for\@citeb:=#2\do
	{\@citea\def\@citea{,}\@ifundefined
	{b@\@citeb}{{\bf ?}\@warning
	{Citation `\@citeb' on page \thepage \space undefined}}
	{\csname b@\@citeb\endcsname}}}{#1}}

\newif\if@cghi
\def\cite{\@cghitrue\@ifnextchar [{\@tempswatrue
	\@citex}{\@tempswafalse\@citex[]}}
\def\citelow{\@cghifalse\@ifnextchar [{\@tempswatrue
	\@citex}{\@tempswafalse\@citex[]}}
\def\@cite#1#2{{$\null^{#1}$\if@tempswa\typeout
	{IJCGA warning: optional citation argument 
	ignored: `#2'} \fi}}

\def\pmb#1{\setbox0=\hbox{#1}
	\kern-.025em\copy0\kern-\wd0
	\kern.05em\copy0\kern-\wd0
	\kern-.025em\raise.0433em\box0}

\def\fnt#1#2{\footnotetext{\kern-.3em
	{$^{\mbox{\scriptsize #1}}$}{#2}}}

\def\runninghead#1#2{\pagestyle{myheadings}
\markboth{{\protect\footnotesize\it{\quad #1}}\hfill}
{\hfill{\protect\footnotesize\it{#2\quad}}}}
\font\tenrm=cmr10
\font\tenit=cmti10 
\font\tenbf=cmbx10
\font\bfit=cmbxti10 at 10pt
\font\ninerm=cmr9

\font\eightrm=cmr8

\def\qed{\hbox{${\vcenter{\vbox{			%HOLLOW SQUARE
   \hrule height 0.4pt\hbox{\vrule width 0.4pt height 6pt
   \kern5pt\vrule width 0.4pt}\hrule height 0.4pt}}}$}}

\renewcommand{\thefootnote}{\fnsymbol{footnote}}	%USE SYMBOLIC FOOTNOTE

\begin{document}

\runninghead{Garc\'{\i}a-Salcedo et al} {BI Cosmologies} 

\normalsize\textlineskip
%\thispagestyle{}
%\setcounter{page}{171}

%\copyrightheading{Vol. , No. (199) }

\vspace*{0.88truein}

%\fpage{171}
\centerline{\bf BORN-INFELD COSMOLOGIES}
\vspace*{0.37truein}
\centerline{\footnotesize Ricardo Garc\'{\i}a-Salcedo and Nora Bret\'on}
\vspace*{0.015truein}
\centerline{\footnotesize\it Depto. de F\'{\i}sica, Cinvestav}
\baselineskip=10pt
\centerline{\footnotesize\it A.P. 14-740, 07000 M\'exico D.F., M\'exico}
\vspace*{0.225truein}
\pub{}

\vspace*{0.21truein}
\abstracts{ We present a model for an inhomogeneous and anisotropic early
universe filled with a nonlinear electromagnetic field of Born-Infeld (BI)
type. The effects of the BI field are compared with the linear case
(Maxwell). Since the curvature invaria nts are well behaved then we
conjecture that our model does not present an initial big bang
singularity. The existence of the BI field modifies the curvature
invariants at $t=0$ as well as sets bounds on the amplitude of the
conformal metric function. 
}{}{} 

%\textlineskip			%) USE THIS MEASUREMENT WHEN THERE IS
%\vspace*{12pt}			%) NO SECTION HEADING

\vspace*{1pt}\textlineskip	%) USE THIS MEASUREMENT WHEN THERE IS
\section{Introduction}	%) A SECTION HEADING
\vspace*{-0.5pt}
\noindent

The advantage in exploring a cosmological model with nonlinear
electrodynamics resides in approaching with a classical model but taking
into account some quantical features. It is well known that nonlinear
electrodynamics (NLE) is the best classical solution to the self-energy
problem of charged particles. The aim of nonlinear electrodynamics of Born
and Infeld (BI) \cite{BI} was to establish a model of a finite classical
field theory without divergences. However, NLE can also be a model of
quantum electrodynamics in the classical limit of high occupation
numbers.  As demonstrated by Stehle and De Baryshe \cite{Stehl}, theories
with Lagrangians similar to the Heisenberg-Euler effective Lagrangians
(the weak field expansion of the Born-Infeld theory is of this form) are
more accurate classical approximations of QED than Maxwell's theory in the
case of fields with high intensities at a fixed frequency. This fact makes
NLE particularly interesting in general relativity, specialy in
cosmological theories, as a simple classical model to explain vacuum
polarization processes- a possible influence on the mechanism of the
evolution of the early universe. The processes called scattering of light
by light can be expressed in terms of a kind of electric and magnetic
permeability of vacuum. From this point of view, the exact solutions of
the Einstein-Born-Infeld (EBI) equations are worth to study since they may
indicate the physical relevance of nonlinear effects in strong
gravitational and strong electromagnetic fields.

The Born-Infeld theory is self-consistent and satisfies all natural
requirements. The BI Lagrangian depends on a constant $b$, that is a
maximum field strenght, and also depends on the invariants of the field.
Recently BI has recover interest in the context of string theory, since
the low energy action Lagrangian corresponding to the objects called {\it
branes} can be written as a term that is BI plus a Weiss-Zumino or
Chern-Simmons term \cite{Tsey}. Furthermore, the BI strings are solitons
which represent strings ending on branes.

In relation to exact solutions of the EBI equations we can mention that
there are exact solutions for spaces possessing one timelike and one
spacelike Killing vector fields. They correspond to spherical static
spacetimes \cite{Pellicer} and also to axisymmetric stationary spacetimes
\cite{Plebas0}.

In the literature cosmological models have been proposed for the
source-free Einstein-Maxwell equations \cite{Melvin}, \cite{Carmeli}.
Charach and Malin \cite{Charach} presented a cosmological model of
Einstein-Rosen with gravitational, electromagnetic and scalar waves, this
work generalized the electromagnetic Gowdy model. In all of these models
the matter fields cause an evolution significantly different from that of
the vacuum model. The interest is to get insight on the coupled
electromagnetic and gravitational radiation at early epochs of the
universe. Since at that epoch the fields must had been very strong, the
nonlinear electromagnetic effects could be of importance. These fields
could originate from processes near the Planck era ($t \sim 10^{-43} s$). 

The spacetime we address in this work possesses two Killing vectors, both
spacelike. It is not immediate to conclude that this spacetime is also
compatible with a BI source. For instance it does not admit a perfect
fluid source together with BI, which is the case in the stationary
axisymmetric spacetime. This is the first atempt, as far as we know, of
exploring BI early universes.  Our aim is to determine to what extent the
BI field can prevent or modify the initial big bang singularity. Spatially
inhomogeneous cosmological models with perfect fluid which do not
originate in an initial singularity have been previously found by
Senovilla \cite{Seno} (see also \cite{Mars}, \cite{lopr}). For the
solutions derived in the present work the initial big bang singularity
appears only as a coordinate singularity, since the invariants do not
diverge at the origin. It turns out that the existence of the BI field set
bounds to the amplitude of the conformal metric function and also affects
quantitatively the curvature invariants at $t=0$. In this work we do not
treat the possible transition of our model to a homogeneous, isotropic
one. In Sec. 2 the Einstein-Born-Infeld equations are established. In Sec. 
3 the solution is presented and analyzed in the linear limit as well as
the asymptotic behavior. In Sec. 3.4 we remark the coordinate
transformations relating the spacetime studied and the Einstein-Rosen
metrics. Concluding remarks are given at the end.

\textheight=7.8truein
\setcounter{footnote}{0}
\renewcommand{\thefootnote}{\alph{footnote}}

\section{Einstein-Born-Infeld Equations}
\noindent

The line element considered corresponds to a $G_2$ spacetime, given by
\begin{equation}
ds^2= \frac{1}{\varphi^2} \{ \frac{dz^2}{h}-\frac{dt^2}{s}+
G[e^{-W}dy^2+e^W( dx+m dy)^2] \},
\label{eq:lineel} 
\end{equation}
where $x$, $y$ are ignorable coordinates (i.e. the spacetime has two
spacelike Killing vectors, $\partial_x, \partial_y$), the metric functions
depend on $z$ and $t$: $\varphi = \varphi(t)$, $h=h(z)$, $s=s(t)$.
$G(z,t)$ and $e^{W}(z,t)$ are separable functions of $z$ and $t$. The
metric (\ref{eq:lineel}) corresponds to a space with gravitational waves
propagating in the $z$ direction, with their wavefronts on the plane
surfaces spanned by the Killing vectors. The function $W$ describes what
is called the ``$+$" polarization of gravitational waves. $G(z,t)$
describes the transverse scale expansion created by the energy density of
the waves. The function $m(z)=m_0+2lz$ is playing the role of unpolarizer
for the propagating gravitational waves that constitute the background.
Making $l \to 0$ the metric recover the form of a space of linearly
polarized gravitational waves (diagonal metric). 

To interpret line element (\ref{eq:lineel}) as cosmological, we refer to
the Einstein-Rosen metric \cite{E-R}, Eq. (\ref{eq:E-R}), Sec. (3.3). The
local behavior of these spacetimes is defined by the gradient of the
metric function $G$, i.e. $G_{, \mu}$ w hich can be spacelike, null or
timelike. The globally spacelike case corresponds to cylindrical
spacetimes. The case of $G_{, \mu}$ being globally null corresponds to the
plane symmetric waves. The other case of $G_{, \mu}$ being globally
timelike as well as cases in which the sign of the gradient can vary from
point to point are used to describe cosmological models and also colliding
gravitational waves \cite{C-Ch}. It must also be remarked that due to the
presence of an Abelian subgroup $G_2$ in the Bianchi models of types
I-VII, the Einstein-Rosen metrics include these models as particular cases
\cite{Tomita}. Also the rotationally invariant Bianchi models of types
VIII and IX belong to the generalized Einstein-Rosen spacetimes
\cite{Feinst}. 

We note that in the case of Einstein-Rosen spacetimes, the interpretation
as gravitational waves propagating in the $z$ direction arises directly
from the equation satisfied by $G$,
\begin {equation}
G_{,tt}-G_{,zz}= 8 \pi G e^{f}(T^a_a-T),
\end{equation}
 
For vacuum or when $T^a_a = T $, $G$ satisfies the usual wave equation. In
our case, the B-I field is not traceless and then $G$ is some propagating
wave with a source (the B-I field). 

Turning back to the spacetime (\ref{eq:lineel}), it is filled with a BI
field with energy-momentum tensor given by: 
\begin {equation}
T_{\mu \nu}= 2 {\cal H}_{,P} P_{\mu \alpha}P^{\alpha}{}_{\nu}-g_{\mu
\nu}[2P{\cal H}_{,P} + Q {\cal H}_{,Q}-{\cal H}],
\label{eq:tem}
\end{equation}
where $\cal H$ is the Born-Infeld structural function and ${\cal H}_{,x}:=
\partial {\cal H}/ \partial x$. $P$ and $Q$ are the invariants associated
with the Born-Infeld field, given by
\begin{equation}
P= \frac{1}{4}P_{\mu \nu}P^{\mu \nu}, \quad Q=\frac{1}{4}P_{\mu \nu}
\check{P}^{\mu \nu},
\end{equation}
$ \check{P}^{\mu \nu}$ denotes the dual tensor of ${P}^{\mu \nu}$,defined
by $\check{P}^{\mu \nu}:= \frac{1}{2} \epsilon^{\mu \nu \alpha
\beta}P_{\alpha \beta}$.  The antisymmetric tensor $P_{\mu \nu}$ is the
generalization of the electromagnetic tensor $F_{\mu \nu}$. The structural
function is constrained to satisfy some physical requirements: (i) the
correspondence to the linear theory $[ {\cal H}(P,Q)=P+O(P^2,Q^2)]$; (ii)
the parity conservation $[{\cal H}(P,Q)={\cal H}(P,-Q)]$; (iii) the
positive definiteness of the energy density (${\cal H}_{,P} >0$) and the
requirement of the timelike nature of the energy flux vector ($P{\cal
H}_{,P}+Q{\cal H}_{,Q}-{\cal H} \ge 0$). Such structural function for the
BI field is given by
\begin{equation}
{\cal H}= b^2- \sqrt{b^4-2b^2P+Q^2},
\label{eq:HBI}
\end{equation}
where $b$ is a maximum field strenght of the BI field.  In the linear
limit, which is obtained by taking $b \to \infty$, then ${\cal H} =P$ and
$P_{\mu \nu}=F_{\mu \nu}$, coinciding with the Maxwell electromagnetism. 
$P_{\mu \nu}$ and $F_{\mu \nu}$ are related through the material or
constitutive equations: 
\begin{equation}
F_{\mu \nu}={\cal H}_{,P} P_{\mu \nu}+ {\cal H}_{,Q} \check{P}_{\mu \nu},
\label{eq:mateqs}
\end{equation}

We have worked out the problem in the null tetrad formalism
\cite{Plebas2}. The function $\cal H$ in Eq. (\ref{eq:HBI}) involves a
square root that complicates the solving of the system of the
Einstein-Born-Infeld equations. To surmount this difficulty, first of all
we have aligned the two non-null components of $P_{\mu \nu}$ along the two
principal directions of the tetrad vectors $e^3$ and $e^4$ (the metric is
type D); second, we parametrize the BI field in terms of the electric and
magnetic fields:  $P_{34}=D$ and $P_{12}=iH$ where $D$ and $H$ stand for
the usual electric displacement and magnetic field intensity,
respectively. Also, we introduce the function $\nu = \nu (D,H)$,
\begin{equation}
\exp{[2 \nu]}= \frac{b^2+D^2}{b^2-H^2},
\label{eq:e2nu}
\end{equation}
If the previous Eq. (\ref{eq:e2nu}) has to make sense, it demands that $b
> H$. From the structural Eqs. (\ref{eq:mateqs}) we obtain the
expressions that clearly manifest the vacuum polarization, in this case
depending on the function $\nu(z,t)$: 
\begin{equation}
D=e^{\nu}E, \quad B=e^{ \nu}H,
\label{eq:DHpol}
\end{equation}

Besides the Einstein field equations we must consider the BI
electrodynamical equations that amount to the closure condition of the
electromagnetic two form, $\omega$,
\begin{equation}
d \omega = d \{(F_{12}+P_{34})e^1 \wedge e^2 +(F_{34}+P_{12})e^3 \wedge
e^4 \}=0,
\end{equation}
This formalism of the Born-Infeld field was introduced by Pleba\~nski et
al\cite{Plebas1}.  We have assumed that $h(z)= \alpha + \beta z+\epsilon
z^2$, where $\epsilon$ can take the values $-1,0,1$. For each value of
$\epsilon$ we obtain different curvatures in the $z$ direction.  For the
metric (\ref{eq:lineel}) and with the alignment chosen the electromagnetic
equations are
\begin{eqnarray}
(E+iH)_{,z}&=&0  ,  \\ 
( \ln (D+iB))_{,t}&=& -2ile^{- \nu}+2(\ln \varphi)_{,t},\
\label{eq:BIeq2}
\end{eqnarray}
Eq.(9) tells us that neither $E$ nor $H$ depend on $z$. Eq.(10) and Eq.
(\ref{eq:nu}), quoted below, lead to infer that neither $D$ nor $B$ depend
on $z$. Therefore, for this spacetime, the BI is a spatially homogeneous
field. The Einstein equations for the BI energy-momentum tensor
(\ref{eq:tem}) amount to
\begin{eqnarray}
\varphi_{,tt}+l^2 \varphi&=&0  ,  \\ 
\frac{\varphi^2}{2}s_{,tt}-2 \varphi \varphi_{,t}s_{,t} +s( 3
\varphi_{,t}^2 +3 l^2 \varphi^2)&=& 2b^2(e^{- \nu}-1),\\
\varphi \varphi_{,t}s_{,t} +s( -3 \varphi_{,t}^2 +l^2 \varphi^2) +\epsilon
\varphi^2 &=& -2b^2(e^{ \nu}-1) \,
\end{eqnarray}

It is a system of equations for the functions $\varphi(t)$, $\nu(t)$ and
$s(t)$, while the rest of the metric functions are $e^{W}=
(s/h)^{\frac{1}{2}}$ and $G(z,t)= (hs)^{\frac{1}{2}}$.  Eq. (11) for
$\varphi (t)$ can be solved immediately, giving $\varphi(t)=A \cos(lt)+B
\sin(lt)$, with $A, B$ constants. Eqs. (12)-(13) can be decoupled and it
is obtained the solution for $\nu(t)$ in terms of $\varphi(t)$ as well as
Eq. (\ref{eq:seq}) fo r $s(t)$: 
\begin{equation}
\nu (t)= \frac{1}{2} \ln (1- \varphi(t)^4),
\label{eq:nu}
\end{equation}

\begin{equation}
\varphi \varphi_{,t}s_{,t} +s( -3 \varphi_{,t}^2 +l^2 \varphi^2) +\epsilon
\varphi^2 = -2b^2(\sqrt{(1- \varphi^4)}-1),
\label{eq:seq}
\end{equation}

In order that $\nu(t)$, given in Eq. (\ref{eq:nu}), be a real function,
$\varphi(t)=A \cos(lt)+B \sin(lt)$ must be such that $\vert \varphi \vert
^4 < 1$. This means that the existence of the BI field restrains the
conformal metric function $\varphi (t)$ in its amplitude because it bounds
the values of the constants $A$ and $B$. Demanding that
\begin{eqnarray}
A <1 & \quad {\rm if} &B=0; \nonumber \\
B <1 & \quad {\rm if} &A=0; \nonumber \\
A^2+B^2 < 1 & \quad {\rm if} &A \ne 0, \quad B \ne 0,
\label{eq:conditions}
\end{eqnarray}
fulfils the requirement $\vert \varphi \vert ^4 < 1$. Putting the NLE
functions, $D(t)$ and $B(t)$, Eqs. (\ref{eq:DHpol}), in terms of
$\varphi(t)$

\begin{eqnarray}
D(t)&=& \varphi^2 \cos[2l \int{(1- \varphi^4)^{- \frac{1}{2}}dt}], \nonumber \\
B(t)&=& - \varphi^2 \sin[2l \int{(1- \varphi^4)^{- \frac{1}{2}}dt}],
\label{eq:D,B}
\end{eqnarray}
with the corresponding coordinate components of the electromagnetic field
$F_{\mu \nu}$,
\begin{eqnarray}
F_{tx}(t)&=&e^{- \nu} \cos{[2l \int{(1- \varphi^4)^{- \frac{1}{2}}dt}]},
\nonumber \\
F_{zy}(t)&=& - e^{\nu} \sin[2l \int{(1- \varphi^4)^{- \frac{1}{2}}dt}],
\end{eqnarray}
The previous expressions show that the BI field modulates the amplitude
and frequency of the electromagnetic field. The fields $B(t)$ and $D(t)$
are shown in Fig. 1. They are oscillating functions depending on the
constant $b$ through the function $e^{\nu }$.

Since the metric is of Petrov type D, the only nonvanishing Weyl scalar is
given by
\begin{equation}
\psi_2= \varphi^2 [\frac{s_{,tt}}{4}- \frac{\epsilon}{2}-2l^2s+i
\frac{3l}{2}s_{,t}],
\end{equation}

This scalar is not real and therefore the magnetic part of the Weyl tensor
does not vanish. We also note that it is homogeneous, i. e. it depends
only on time.  The invariants are given by $I=3 \psi_2^2, \quad J=-
\psi_2^3$; from their expressions it is clear that they share (if they
exist) the singularities of $\psi_2$. On its turn $\psi_2$ depends on the
behaviour of the functions $\varphi (t)$ and $s(t)$. $\varphi (t)$ is a
periodic function while $s(t)$, governed by Eq. (\ref{eq:seq}), carries
much of NLEBI information. In the next section we analyze the behavior of
$s(t)$.

\section{The BI solution and the linear case}
\noindent

The solution to Eq. (\ref{eq:seq}) for $s(t)$, in terms of $\varphi$ is
given by
\begin{equation}
s(t)= \varphi^3 \varphi_{,t} \{ c_1 - \int{ \frac{2b^2 ( \sqrt{1-
\varphi^4} -1)+ \epsilon \varphi^2}{\varphi^4 \varphi_{,t}^2}dt} \},
\label{eq:ssol}
\end{equation}
where $c_1$ is an integration constant related to initial or boundary
conditions for vacuum. It is interesting now to analyze graphics of the
function $s(t)$.  Fig. 2 shows $s(t)$ for several values of the BI
constant $b$. For values of $b > 1$ the function $s(t)$ becomes positive
for all the range and reduces its oscillations; for larger $b$ t he
maximums become greater. The transverse scale expansion $G(z,t)=
\sqrt{h(z)s(t)}$ is an inhomogeneous function expanding in the $z$
direction, it is shown in Fig.3.

Fig. 4 shows the plots of the real and imaginary parts of $\psi_2$; they
give us qualitative information with respect to the singularities of the
invariants $I$ and $J$. Both Re$\psi_2$ and Im$\psi_2$ are continuous
functions that do not present infinities at all. 

In Fig. 5, the function $s(t)$ displays different behaviours at $t=0$
depending if $\varphi=B \sin(lt)$ or $\varphi= A \cos(lt)$. In the former
case $s(0)=0$ and consequently there appears a coordinate singularity at
$t=0$; however the corresponding Weyl scalar is not singular there.  We
can see from Fig. 5 that this class of solutions admits both behaviors at
$t=0$, with coordinate singularity or without it. To a linear combination
$\varphi= A \cos(lt) + B \sin(lt)$ corresponds $s(0) \ne 0$.

%*************************************************************

\subsection{Linear limit}
 
In order to compare the effects of the BI field with the Maxwell case, we
compute the linear limit. The Maxwell electrodynamics is recovered when $b
\to \infty$. The BI Eqs. (9) and (\ref{eq:BIeq2}) in the linear limit
correspond to
\begin{equation}
[\ln (D+iB)]_{,t}-2(\ln \varphi)_{,t}=0,
\end{equation}
whose solution is
\begin{equation}
D=E=C \cos (2Alt), \quad B=H=C \sin(2Alt),
\end{equation}
where $A$ and $C$ are constants. In terms of the coordinate components of
the electromagnetic field the expressions are
\begin{equation}
F_{tx} = C \cos (2lt), \quad F_{zy}= - C \sin(2lt),
\end{equation}

This electromagnetic field is homogeneous of intensity $C^2=E^2+H^2$.  In
accordance with the parametrization (\ref{eq:e2nu}), for $b \to \infty$,
Eqs. (12)-(13) become
\begin{eqnarray}
\frac{\varphi^2}{2}s_{,tt}-2 \varphi \varphi_{,t}s_{,t} +s( 3
\varphi_{,t}^2 +3 l^2 \varphi^2)&=& -C^2,\nonumber\\ \varphi
\varphi_{,t}s_{,t} +s( -3 \varphi_{,t}^2 +l^2 \varphi^2) +\epsilon
\varphi^2 &=& -C^2 \,,
\end{eqnarray}

Solving for $s(t)$ in terms of $\varphi(t)$ we have
\begin{equation}
s(t)= \varphi^3 \varphi_{,t}[c_1- \int{\frac{C^2 + \epsilon
\varphi^2}{\varphi_{,t}^2 \varphi^4}dt}],
\label{eq:sli}
\end{equation}
expression that also can be obtained from Eq.(\ref{eq:ssol}) taking the
limit $b \to \infty$,
$$ \lim_{b \to \infty} 2b^2 (\sqrt{1- \varphi^4}-1) = \lim_{b \to \infty}
2 b^2 (e^{\nu}-1)= C^2.$$
Fig. 6 shows the comparison between $s(t)$ corresponding to the linear
electromagnetic field, Eq. (\ref{eq:sli}), and $s(t)$ of the BI field, Eq.
(\ref{eq:ssol}). The presence of BI field smooths the function $s(t)$
respect of the linear version. Making $C=0$ we get the vacuum limit of the
spacetime (\ref{eq:lineel}), by this meaning that only gravitational waves
remain.

Comparing Eq.(\ref{eq:ssol}) for $s(t)$ corresponding to the BI field with
Eq.(\ref{eq:sli}) for the linear case we see that the first term in the
integral is directly associated with the NLEBI effects. To show the
dependence on $b$ we write $s(t)$ as
\begin{equation}
s(t)= \varphi^3 \varphi_{,t}[c_1- \int{\frac{f(b) + \epsilon
\varphi^2}{\varphi_{,t}^2 \varphi^4}dt}],
\end{equation}
where $f(b)=2b^2(e^{ \nu}-1)=2b^2 \{ \sqrt{\frac{b^2+D^2}{b^2-H^2}} -1
\}$, in accordance with the requirement in Eq. (\ref{eq:e2nu}),the range
of $b$ is $b >H$.  If one plots $f(b)$ vs. $b$, it can be seen that the
NLEBI effects are sensible for values of $b$ near $H$, while for larger
$b's$, generically it tends to the constant linear Maxwell field
$C^2=E^2+H^2$. 

%************************************************************

\subsection{Behavior at early times}

For $t << 1$ it is reasonable to approximate $\varphi(t)$ as  
\begin{equation}
\varphi(t)= A \cos (lt) +B \sin (lt) \approx A +Blt,
\end{equation}
Since $ \varphi =$ const. implies $s(t)=0$, we take $\varphi \approx t$
along with the restrictions (\ref{eq:conditions}) imposed by the BI field.
The constant $l$ is suposed not to be large since we are in a low
frequency regime. To solve Eq. (\ref{eq:ssol}) we also approximate $
\sqrt{1- \varphi^4} \approx 1 - \frac{\varphi^4}{2}$, obtaining
\begin{equation}
s(t) \approx \epsilon t^2+c_1 t^3+b^2[t^4+ O(t^5)], 
\label{eq:searly}
\end{equation}
predominance of one term over the others depend on the relative values
between $c_1$ and $b$. The term in brackets represents the BI contribution
at early times. It is expected $b$ to be large if the electromagnetic
field is strong, then the BI contributi on can be of importance.  The
first term corresponds to the contribution of the space curvature,
determined by the value of $\epsilon, (1,-1,0).$ The Weyl scalar $\psi_2$
approaches $t=0$ as
\begin{equation}
\psi_2= \frac{3c_1}{2}t^3+(3b^2-2l^2 \epsilon)t^4 + O(t^5)+ 
i (3 \epsilon lt^3+ \frac{9c_1 l}{2}t^4+ O(t^5) ).
\end{equation}

The contribution of the BI field goes as $t^4$ and higher orders in $t$.
In the magnetic part of $\psi_2$ the contribution is not important at very
early times since comes from the term with $\epsilon=-1,0,1$ and goes as
$t^3$.

We now mention the case $l=0$, which corresponds to a spacetime of
propagating linearly polarized gravitational plane waves (diagonal
metric). Eq. (11)  for $\varphi(t)$ become $\varphi_{,tt}=0$, with
solution $A_0 t +B_0$, $A_0, B_0$ being constants. The refore, the
behavior of $\varphi$ in the case $l=0$ resembles the one for $t << 1$.
Consequently, we assert that this spacetime approaches the origin in time
as propagating linearly polarized waves.

%***********************************************

\subsection{Behavior  at $t >> 1$ }

To complete the previous analysis, the question arises as how does the
model evolve at $t>>1$, or which is the asymptotic behavior of the
solution at $t >>1$, i. e. for times when the universe is already a
causally connected one. 

The dynamics of our model is driven by the function $\varphi (t)$. For
$t>>1$ both $\varphi$ and $\varphi_{,t}$ become rapidly oscillating
periodic functions that can be approximated by a constant. With this, the
metric function $s(t)$ (Eq. (\ref{eq:ssol}) becomes a linear function on
$t$. Absorbing constants and transforming $\sqrt{K_1-K_2 t} \to K_2 T/2$
($K_1, K_2$ being constants), the line element can be written as
\begin{equation}
ds^2= \frac{dz^2}{h(z)}-dT^2+ h(z)dy^2+ \frac{K_2 T}{2}( dx+m dy)^2,
\label{eq:asymp} 
\end{equation}

While the nonlinear electromagnetic field for $t>>1$, from Eq.
(\ref{eq:nu}), we see that $\nu(t) $ goes to a constant. Then, from Eqs.
(\ref{eq:D,B}) for the fields $D(t), B(t)$, we recover the Maxwell case.

%*************************************************************

\subsection{Relation to Einstein-Rosen Universes}

By means of a coordinate transformation, the metric (\ref{eq:lineel}) can
be put in the form of an Einstein-Rosen line element \cite{E-R}
\begin{equation}
ds^2= \frac{1}{\varphi ^2} \{ dZ^2-dT^2 +  G[e^{-W}dy^2+e^W( dx+m dy)^2] \}.
\label{eq:E-R}
\end{equation}

The coordinate transformation for $z \to Z$ depends on the value of
$\epsilon$ as follows
\begin{eqnarray}
z= (\frac{\beta^2}{4}- \alpha)^{\frac{1}{2}} \cosh Z - \frac{\beta}{2},&&
\quad \epsilon=1, \\ z= (\frac{\beta^2}{4}+ \alpha)^{\frac{1}{2}} \sin Z +
\frac{\beta}{2}, && \quad \epsilon=-1, \\ z= \frac{Z^2 \beta}{4}-
\frac{\alpha}{\beta}, && \quad \epsilon=0,
\end{eqnarray}

We also note that for $\epsilon=-1$, the functions in terms of the spatial
coordinate $z$ become periodic, then the topology of the universe in this
case can be a closed one.  The coordinate transformation for $t$, $
\frac{dt^2}{s} \to dT^2$, is not a simple one in the general case due to
the form of $s(t)$, Eq. (\ref{eq:ssol}). It involves elliptic integrals
which lead to Jacobi family of elliptic functions \cite{Breton} and there
is no analytical expression in terms of elementary functions. However, for
particular cases such transformation can be given in a simple form.  An
example is for early times. In this case the relation between $t$ and $T$
is $exp(\sqrt{\epsilon} T)= t^{-2}(2 \epsilon t +c_1 t^2+2t \sqrt{\epsilon
(\epsilon+c_1 t +b^2 t^2)})$.

Another example is the case $b=0$ (vacuum) and $ \epsilon=0$.  In this
case the line element (\ref{eq:E-R}) takes the form
\begin{equation}
ds^2= \frac{1}{\varphi ^2}  (dZ^2-dT^2) +  TZ[e^{-W}dy^2+e^W( dx+m dy)^2],
\label{eq:Gow0}
\end{equation}

Gowdy \cite{Gowdy} constructed exact vacuum solutions of the Einstein
field equations which represent inhomogeneous closed universes. These
models possesses compact spacelike hypersurfaces as well as $G_2$
invariance. The line element (\ref{eq:Gow0}) can be written as a Gowdy
model with thre-torus topology performing the coordinate transformation
\begin{equation}
G=TZ \to  \xi, \quad  \frac{Z^2}{2}+ \frac{T^2}{2} \to \zeta,
\end{equation}
then one obtains the three-torus Gowdy line element
\begin{equation}
ds^2= \frac{\varphi ^{-2}}{2 \sqrt{\zeta^2 -\xi^2}} (d \zeta^2-d \xi^2) +
\xi [e^{-W}dy^2+e^W( dx+m dy)^2],
\label{eq:Gow1}
\end{equation}

If one arrives to a Gowdy model for vacuum, one can speculate about that a
geometry so simple as Gowdy does not admit a Born-Infeld field. 

\section{Concluding Remarks}
\noindent

In this work we have presented a family of solutions to the
Einstein-Born-Infeld equations for a space-time which is a $G_2$
cosmological model. The spacetime describes propagating gravitational
plane waves coupled with a nonlinear spatially homogeneous e
lectromagnetic field of the Born-Infeld type. We also present the limit in
which the BI field become Maxwell electromagnetism. 

Some results are:

The presence of an initial coordinate singularity depends on the choice of
the conformal metric function $\varphi(t)=A \cos(lt)+ B \sin(lt)$, for $B
\ne 0, A = 0$ the solution exhibites initial singularity while if $B = 0,
A \ne 0$ there is no such coordi nate singularity. However, from the
smooth behavior of the Weyl scalar, $\psi_2$, we guess that there are no
singularities at all.

The presence of the BI field sets bounds on the amplitude of the periodic
function $\varphi (t)$. In this sense BI modifies the global expansion of
the spacetime. If compared with the effect of linear electrodynamics, the
presence of the BI field smooths the metric function $s(t)$ (see Fig. 6).
Since $s(t)$ is involved in the expression of the scale expansion and of
the Weyl scalars, then BI field smooths the curvature.  The BI field also
modifies quantitatively the curvature at $t=0$, the effect being more
sensible when the BI parameter $b$ approaches the magnitude of a critic
magnetic field intensity $H$. We conjecture that a less restrictive
spacetime (not type D) shall permit the existence of an inhomogeneous
nonlinear electromagnetic field, whose consequences on the space curvature
could be more drastic.

For early times, $t \to 0$, the spacetime approaches a space of
unpolarized gravitational waves.  For $t >> 1$ the spacetime becomes an
inhomogeneous anisotropic spacetime with a homogeneous Maxwell field. 

For vacuum one obtains a Gowdy model of three-torus topology.  
 
It should be of interest to classify the spacetime in different regions
according to the sign of the gradient of the function $G(z,t)$ and see the
distinct interpretations of the found solution according to each region in
the $(z,t)$ plane. 

Generalizing this solution to include a scalar field and to investigate
the coupled effect with nonlinear electrodynamics for early universes
could also lead to interesting results. 

\nonumsection{Acknowledgements}
\noindent
 
We gratefully ackowledge to the anonymous referee whose opinions help to
improve this work. Partially supported by CONACyT (M\'exico), project No.
32086-E.

\nonumsection{References}

\newpage 

%%%%%%%%%%%%%%%%%%%%%%%%%%%%%%%%
\vskip-1cm
\begin{center}
\hspace*{0truecm}
\epsfxsize=10truecm
\epsfbox{bretonfig1.eps}
\end{center}
\vskip-0.5cm
\centerline{
\begin{minipage}{16truecm}
\baselineskip14pt
\smallskip\bigskip
{\small 
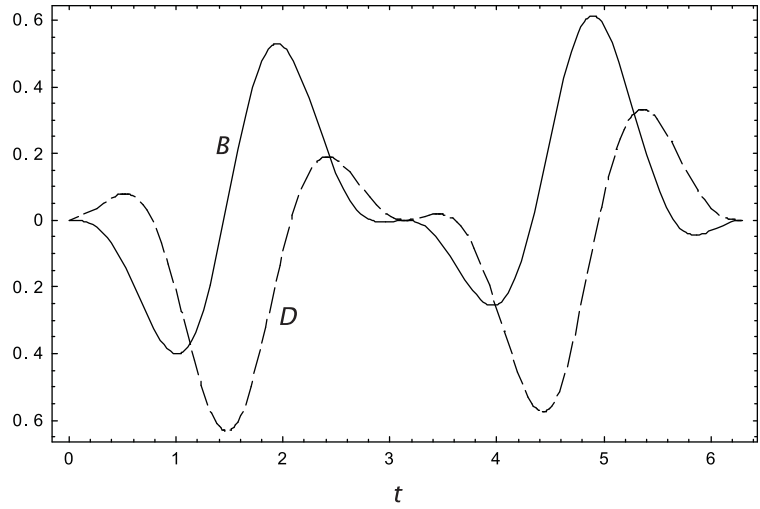
Fig.1. These are plots of the nonlinear electromagnetic fields, electric
displacement, $D(t)$, and magnetic induction, $B(t)$. The corresponding
conformal metric function is $\varphi= 0.8 \sin (t)$.
}
\end{minipage}}

\vskip1cm 

%%%%%%%%%%%%%%%%%%%%%%%%%%%%%%%%
\vskip-1cm
\begin{center}
\hspace*{0truecm}
\epsfxsize=10truecm
\epsfbox{bretonfig2.eps}
\end{center}
\vskip-0.5cm
\centerline{
\begin{minipage}{16truecm}
\baselineskip14pt
\smallskip\bigskip
{\small 
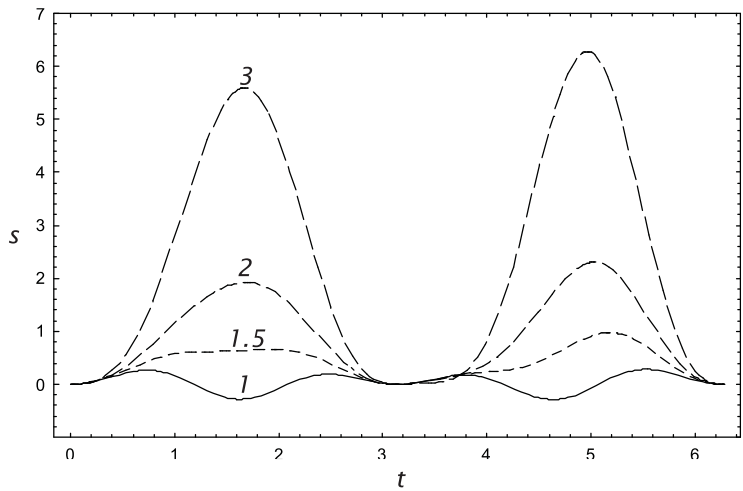
Fig.2.  It is displayed the metric function $s(t)$ for different values of
the BI constant $b$, $b=1, 1.5, 2, 3$.  For greater values of $b$ the
maximums are higher but the function preserves the shape of $b > 2$.
These plots correspond to $\varphi = 0.  8 \sin(t)$, $c_1 =1$ and
$\epsilon =1$.
}
\end{minipage}}

\vskip1cm 

%%%%%%%%%%%%%%%%%%%%%%%%%%%%%%%%
\vskip-1cm
\begin{center}
\hspace*{0truecm}
\epsfxsize=10truecm
\epsfbox{bretonfig3.eps}
\end{center}
\vskip-0.5cm
\centerline{
\begin{minipage}{16truecm}
\baselineskip14pt
\smallskip\bigskip
{\small 
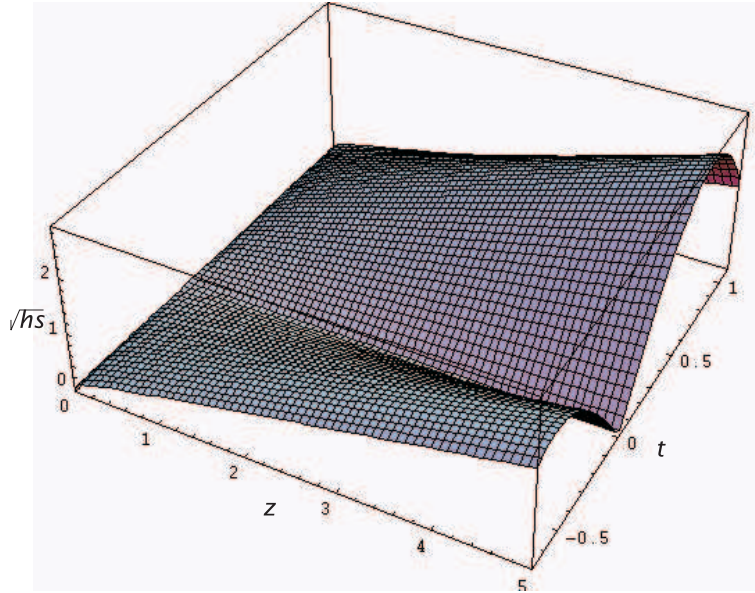
Fig.3. It is shown the transverse scale expansion $G(z,t)= \sqrt{h(z)
s(t)}$. It is increasing in $z$ and oscillating in $t$. For this plot the
values of the constants are $\epsilon =1, \beta =1, \alpha =0, c_1 =1,
b=1$.
}
\end{minipage}}

\vskip1cm 

%%%%%%%%%%%%%%%%%%%%%%%%%%%%%%%%
\vskip-1cm
\begin{center}
\hspace*{0truecm}
\epsfxsize=10truecm
\epsfbox{bretonfig4.eps}
\end{center}
\vskip-0.5cm
\centerline{
\begin{minipage}{16truecm}
\baselineskip14pt
\smallskip\bigskip
{\small 
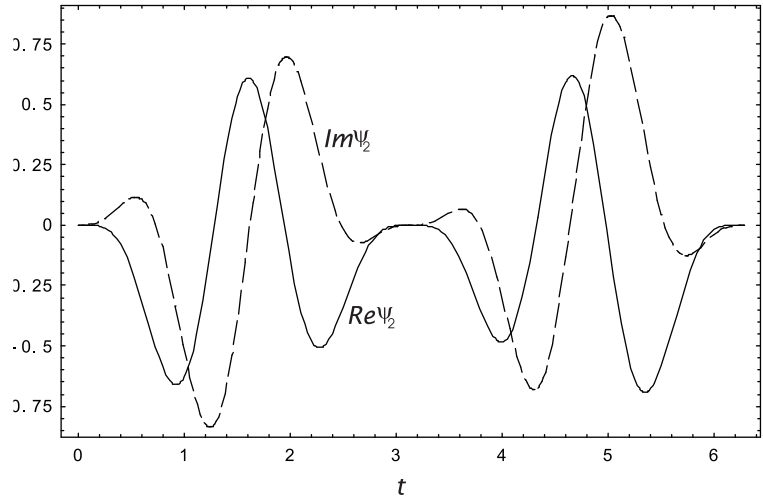
Fig.4. The continuous plot corresponds to Re$ \psi_2$ and the dashed one
to Im$\psi_2$. Both are continuous functions, on this basis we guess that
the spacetime has no singularities. These graphics correspond to $s(t)$ in
Fig.2 with $b=1$. 
}
\end{minipage}}

\vskip1cm 

%%%%%%%%%%%%%%%%%%%%%%%%%%%%%%%%
\vskip-1cm
\begin{center}
\hspace*{0truecm}
\epsfxsize=10truecm
\epsfbox{bretonfig5.eps}
\end{center}
\vskip-0.5cm
\centerline{
\begin{minipage}{16truecm}
\baselineskip14pt
\smallskip\bigskip
{\small 
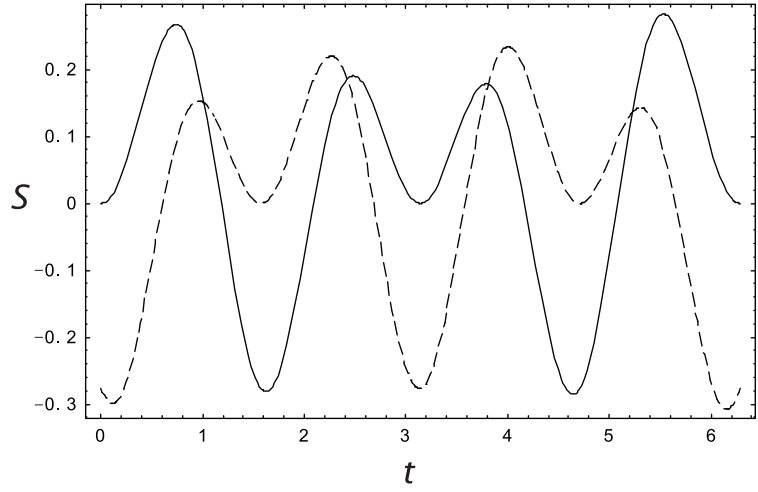
Fig.5. The function $s(t)$ displays different behaviour at $t=0$ depending
if $\varphi= 0.8 \sin(t)$ (continuous curve) or $\varphi= 0.8 \cos(t)$
(dashed curve). 
}
\end{minipage}}

\vskip1cm 

%%%%%%%%%%%%%%%%%%%%%%%%%%%%%%%%
\vskip-1cm
\begin{center}
\hspace*{0truecm}
\epsfxsize=10truecm
\epsfbox{bretonfig6.eps}
\end{center}
\vskip-0.5cm
\centerline{
\begin{minipage}{16truecm}
\baselineskip14pt
\smallskip\bigskip
{\small 
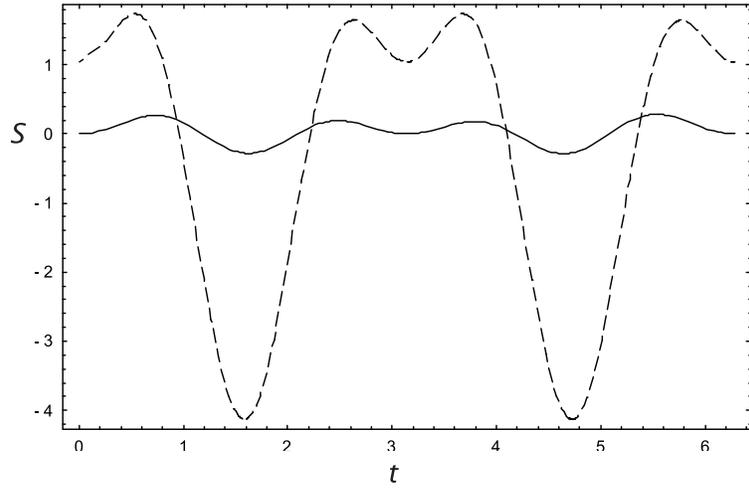
Fig.6. The function $s(t)$ for the linear limit corresponds to the dashed
curve, the continuous line is $s(t)$ in the presence of BI field with
$b=1$. The constant $c_1=1$. $s(t)$ of BI field is smoother than the one
corresponding to linear electromagnetic field. The conformal metric
function considered is $\varphi = 0.8 \sin (t)$.
}
\end{minipage}}

\end{document}